\documentclass[aps,prb,twocolumn]{revtex4-2}

\usepackage{amsmath}
\usepackage{amssymb}
\usepackage{bm}
\usepackage{amsfonts}
\usepackage{graphicx}
\usepackage{booktabs}
\usepackage{xcolor}
\usepackage{hyperref}
\setcitestyle{super,open={},close={}}

\begin{document}

\title{Trimer Thouless Pump: Topology, Symmetries, and Multigap Structure}

\author{Rittwik Chatterjee}
\email{rittwikchatterjee@gmail.com} 
\thanks{ORCID: \href{https://orcid.org/0009-0004-6546-6909}{0009-0004-6546-6909}}

\affiliation{Department of Physics, Jadavpur University,\\
188 Raja Subodh Chandra Mallik Road, Kolkata 700032, India}

\begin{abstract}
Topological charge pumping is paradigmatically understood through two-band systems such as the Rice-Mele model, which are intrinsically restricted to a single independent pumping channel. In this work, we introduce the Trimer Thouless Pump (TTP), a minimal three-band generalization that exhibits genuinely multigap topological transport. By subjecting a one-dimensional three-site lattice to cyclic adiabatic modulations of its hopping amplitudes and antisymmetric onsite potentials, we explore a topological regime characterized by two independent bulk gaps. We show that the quantized charge transport is driven by highly localized Berry curvature hotspots corresponding to effective two-level Dirac avoided crossings on the parameter torus. Crucially, we demonstrate that the middle energy band acts as a geometric mediator: it facilitates the exchange of quantized Berry flux between the upper and lower bands while maintaining a net zero Chern number itself. This bulk topology is corroborated by the spectral flow of boundary-localized edge states traversing multiple gaps. Furthermore, we map the topological phase diagram as a function of central-site detuning, illustrating a band-selective transfer of topological invariants across discrete phase transitions. Finally, we propose a concrete experimental protocol to realize the TTP and observe its multigap charge transport using ultracold atoms in phase-controlled optical superlattices.
\end{abstract}


\maketitle

\section{Introduction}

Topological charge pumping~\cite{wang2013topological} provides a transparent setting in which geometric properties of Bloch bands lead to quantized physical transport. In a Thouless pump~\cite{PhysRevB.27.6083,cooper2019topological,ozawa2019topological}, a one-dimensional insulator is subjected to a slow, periodic modulation of its Hamiltonian. Although the system remains strictly one-dimensional at every instant, a full modulation cycle transports an integer number of particles across the sample. This quantized response is robust to disorder and microscopic details and reflects global topological features rather than the specifics of the modulation protocol.

Adiabatic pumping~\cite{greschner2020topological} introduces an additional control parameter, typically denoted by $\varphi$, in addition to the crystal momentum $k$. The instantaneous Bloch Hamiltonian $H(k,\varphi)$ is therefore defined on a two-dimensional parameter space with the topology of a torus. On this $(k,\varphi)$ torus one may define a Berry curvature~\cite{berry1984quantal} $F_n(k,\varphi)$ for each band $n$, whose integral yields the band Chern number and equals the quantized particle transport per cycle. The Chern number can be inferred from the winding of hybrid Wannier centers, spectral flow of edge states, or direct center-of-mass shifts in cold-atom and photonic experiments~\cite{nakajima2016topological,lohse2016thouless}. For numerical computations, the Berry curvature on the discretized parameter space is efficiently evaluated via the gauge-invariant method of Fukui, Hatsugai, and Suzuki~\cite{Fukui2005}.

A canonical platform for realizing this physics is the Rice--Mele (RM) model~\cite{rice1982elementary}, a dimerized tight-binding chain originally introduced to describe soliton formation in polyacetylene. Its two-band structure and single driving parameter yield a single bulk gap whose topological winding produces quantized charge transport. Because of its analytical tractability and experimental accessibility, the RM model underlies many modern implementations of Thouless pumps. Recent advances in cold-atom and photonic systems now enable the engineering of more complex lattice geometries, motivating an exploration of multiband pumps that go beyond the two-band RM framework~\cite{Mostaan2022,4d5s-n4gn}. While the original RM model is restricted to a single pumping channel defined by two bands~\cite{PhysRevB.27.6083, rice1982elementary}, modern multi-band platforms allow for the observation of non-Abelian topological effects~\cite{You2022} and the redistribution of Berry flux across multiple bulk gaps, offering a richer landscape for topological state engineering.

The exploration of these multiband systems draws upon broader investigations into topological phases under periodic modulation. While the adiabatic TTP protocol relies on slow, periodic modulation to transport charge, our previous work on extended Su--Schrieffer--Heeger (SSH)~\cite{su1979solitons,xie2019topological,zhou2023exploring} models has established that even in non-adiabatic, Floquet-driven regimes, higher winding numbers provide a robust topological signature~\cite{chatterjee2025floquet}. The contrast between the adiabatic limit explored here and the quenched Floquet dynamics examined previously underscores that topological protection in lattices with nontrivial unit cells is a pervasive feature of the underlying band structure and symmetry-anchored avoided crossings, rather than a phenomenon restricted to a single driving limit.

Nevertheless, the standard two-band RM framework is intrinsically restricted to a single independent pumping channel and cannot capture multiband features such as gap competition, redistribution of Berry curvature between bands, or multiple gap-closing pathways. Such phenomena naturally arise in lattices with more than two sublattices per unit cell and are increasingly relevant in experimental platforms capable of realizing multi-site unit cells and dynamic control of sublattice energies.

A trimer chain~\cite{verma2024bulk,ghuneim2024topological,anastasiadis2022bulk}, which we refer to as the \emph{Trimer Thouless Pump} (TTP), provides the next minimal generalization of the Rice--Mele model. Introducing a three-site unit cell yields three energy bands and two independent band gaps~\cite{anastasiadis2022bulk}, enabling qualitatively richer pumping behavior. The presence of three sublattices allows for asymmetric onsite modulations and intracell/intercell hopping patterns that vary throughout the pumping cycle. These ingredients permit topological phase transitions driven by distinct gap closings and open the possibility of redistributing Chern numbers between bands. As a result, trimer pumps can exhibit nontrivial topology in multiple gaps, band-dependent pumping channels, and edge states whose evolution differs between upper and lower gaps---phenomena that lie beyond any two-band description. In this work, we develop and analyze a trimer-based Thouless pump with $\varphi$-dependent intracell hoppings and sublattice potentials. We compute the Berry curvature on the $(k,\varphi)$ torus using the gauge-invariant discretization of Ref.~\onlinecite{Fukui2005}, determine the band-resolved Chern numbers, and corroborate these results by examining spectral flow and localization in open chains. We also discuss possible experimental realizations~\cite{zilberberg2018photonic} in cold-atom, photonic, and topoelectrical platforms. The trimer model thus serves as a minimal setting in which the richer structure of multiband Thouless pumping becomes transparent.

\section{Trimer Hamiltonian and Symmetry Structure}

\begin{figure}[h!]
    \centering
    \def\svgwidth{\columnwidth}
    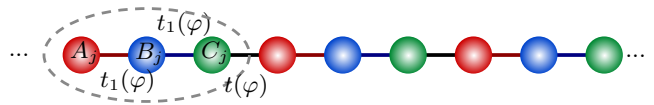
    \caption{Schematic of the Trimer Thouless Pump (TTP).}
    \label{fig:trimer_schematic}
\end{figure}

We consider a one-dimensional tight-binding lattice with a three-site unit cell labeled $A$, $B$, and $C$, as illustrated in Fig.~\ref{fig:trimer_schematic}. The pump parameter $\varphi$ modulates the intracell hoppings and sublattice potentials over a cycle of period $2\pi$. Throughout this work, we set the energy scale by fixing $t_0 = 1$, where $t_0$ represents the average hopping amplitude. The parameter $\delta_t$ controls the strength of the hopping modulation, while $M$ sets the amplitude of the onsite potential modulation. These parameters allow us to interpolate between weakly and strongly modulated pumping regimes. For clarity, we refer to this three-site generalization of the Rice--Mele pump as the \emph{Trimer Thouless Pump} (TTP). The unit cell contains three inequivalent sublattices and supports two intracell hoppings and a single intercell hopping, yielding a three-band structure with two distinct bulk gaps. These features go beyond the minimal two-band Rice--Mele model and are essential for capturing multiband topological pumping.

\subsection{Real-space Hamiltonian}

The second-quantized Hamiltonian for a chain of $N$ unit cells is
\begin{equation}
\begin{aligned}
H(\varphi) = \sum_{j=1}^{N} \Big[
&\, t_1(\varphi)\,(a_j^\dagger b_j + b_j^\dagger c_j)
   + t(\varphi)\, c_j^\dagger a_{j+1}
   + \text{H.c.} \\
&\, + V_A(\varphi)\, a_j^\dagger a_j
   + V_C(\varphi)\, c_j^\dagger c_j
\Big].
\end{aligned}
\label{eq:H_real}
\end{equation}
where $a_j$, $b_j$, and $c_j$ annihilate particles on sublattices
$A$, $B$, and $C$ of unit cell $j$. The modulated hoppings are defined as
\begin{equation}
    t_1(\varphi) = t_0 + \delta_t \cos\varphi, \qquad
    t(\varphi)  = t_0 - \delta_t \cos\varphi,
    \label{eq:tphi}
\end{equation}
producing an exchange of strong and weak bonds analogous to the
Rice--Mele pumping protocol but generalized to a trimer lattice. The
onsite potentials vary as
\begin{equation}
    V_A(\varphi) = M \sin\varphi, \qquad
    V_C(\varphi) = -M \sin\varphi,
    \label{eq:Vphi}
\end{equation}
with $B$ chosen as the energy reference. These specific functional 
forms are chosen for conceptual clarity and experimental relevance:
they preserve the cyclic structure of the RM pump while allowing
independent control over hopping modulation and sublattice asymmetry,
both of which are essential for accessing genuine multiband topological
effects.

\subsection{Bloch Hamiltonian}

Assuming periodic boundary conditions and inserting the Bloch
expansion $a_j = \frac{1}{\sqrt{N}} \sum_k a_k e^{ikj}$ (and analogously for $b_j$, $c_j$), the
Hamiltonian reduces to a $3\times 3$ Bloch Hamiltonian,
\begin{equation}
    H(k,\varphi) =
    \begin{pmatrix}
        V_A(\varphi)      & t_1(\varphi)        & t(\varphi)e^{-ik} \\
        t_1(\varphi)      & 0                   & t_1(\varphi)      \\
        t(\varphi)e^{ik}  & t_1(\varphi)        & V_C(\varphi)
    \end{pmatrix}.
    \label{eq:Hkphi}
\end{equation}
Here $t_1(\varphi)$ couples $A$ to $B$ and $B$ to $C$ within a unit cell,
while $t(\varphi)$ couples site $C$ of unit cell $j$ to site $A$ of the
neighboring cell $j+1$. Diagonalization of Eq.~\eqref{eq:Hkphi} yields
three instantaneous energy bands $E_n(k,\varphi)$ and corresponding
eigenstates $|u_n(k,\varphi)\rangle$. Because the pair $(k,\varphi)$
spans a two-dimensional torus, these eigenstates support Berry curvature
and band Chern numbers that characterize topological pumping.

\subsection{Symmetry Analysis}

The symmetry properties of the TTP play a central role in determining its multiband topology. At special values of the pump phase $\varphi$, the trimer Hamiltonian acquires instantaneous inversion symmetry, while the full pumping cycle breaks time-reversal symmetry and generates nonvanishing Chern numbers. In addition, the three-band structure generically breaks simple chiral symmetry, placing the model outside the symmetry classes that protect the two-band Rice--Mele topological pump.

\subsubsection{Inversion symmetry at $\sin\varphi=0$}

When $\sin\varphi=0$, the onsite potentials vanish, $V_A(\varphi)=V_C(\varphi)=0$, and the Hamiltonian in Eq.~\eqref{eq:Hkphi} becomes inversion symmetric with respect to the $B$ site. In the sublattice basis $(A,B,C)$, the inversion operator is represented by
\begin{equation}
P =
\begin{pmatrix}
0 & 0 & 1 \\
0 & 1 & 0 \\
1 & 0 & 0
\end{pmatrix},
\qquad
P^2 = \mathbb{I},
\end{equation}
and satisfies
\begin{equation}
P\, H(k,\varphi)\, P^{-1} = H(-k,\varphi)
\qquad \text{for } \sin\varphi=0.
\end{equation}
At these symmetry points, the instantaneous Zak phase of each band is quantized to $0$ or $\pi$ (mod $2\pi$), even though the full pumping protocol does not preserve inversion symmetry at all values of $\varphi$. These special values of $\varphi$ act as topological anchors that constrain the location of band inversions and strongly influence the redistribution of Berry curvature across the three bands.

\subsubsection{Instantaneous versus global time-reversal symmetry}

Because all hopping amplitudes and onsite potentials are real numbers (with no Peierls phases or external magnetic flux), the system possesses spinless time-reversal symmetry at each instantaneous phase $\varphi$. The time-reversal operator $\mathcal{T}$ is represented purely by complex conjugation, $\mathcal{T} = K$, where $K$ maps $i \to -i$. The instantaneous Bloch Hamiltonian satisfies
\begin{equation}
\mathcal{T} H(k,\varphi)\mathcal{T}^{-1} = H(-k,\varphi).
\label{eq:TR_instantaneous}
\end{equation}
Thus, for each fixed snapshot of $\varphi$, the system lies in the spinless time-reversal symmetry class AI. 

However, on the full two-dimensional $(k,\varphi)$ parameter torus, full time-reversal symmetry would require invariance under $(k,\varphi) \mapsto (-k,-\varphi)$ combined with complex conjugation. Our driving protocol is not invariant under this transformation because the parameter vector $(t_1(\varphi), t(\varphi), V_A(\varphi), V_C(\varphi))$ traces a directed, non-invariant loop in parameter space as $\varphi$ advances from $0$ to $2\pi$ (since $V_A(-\varphi) = -V_A(\varphi) \neq V_A(\varphi)$ when $\sin\varphi \neq 0$). 

Consequently, the effective two-dimensional problem on the $(k,\varphi)$ torus belongs to Wigner--Dyson symmetry class A\cite{altland1997nonstandard,schnyder2008classification}. The Berry curvature $F_n(k,\varphi)$ is generically nonzero, and the band Chern numbers $C_n$ are nonvanishing.

\subsubsection{Absence of a simple chiral symmetry and multigap topology}

The Rice--Mele model can exhibit an effective chiral symmetry when onsite potentials vanish, and its topology is then protected by sublattice parity. In contrast, the trimer Hamiltonian generically breaks any simple, $k$-independent chiral symmetry. Even at the inversion-symmetric values $\varphi=0,\pi$ with $V_A=V_C=0$, the presence of three inequivalent sublattices and the pattern of intracell and intercell hoppings prevent $H(k,\varphi)$ from being brought into a strictly off-diagonal bipartite form by a $k$-independent unitary transformation. As a result, the topological properties of the TTP are not protected by chiral symmetry and instead are encoded in the global phase of the adiabatic pumping cycle.

\begin{figure}[t]
    \centering
    \includegraphics[width=\linewidth]{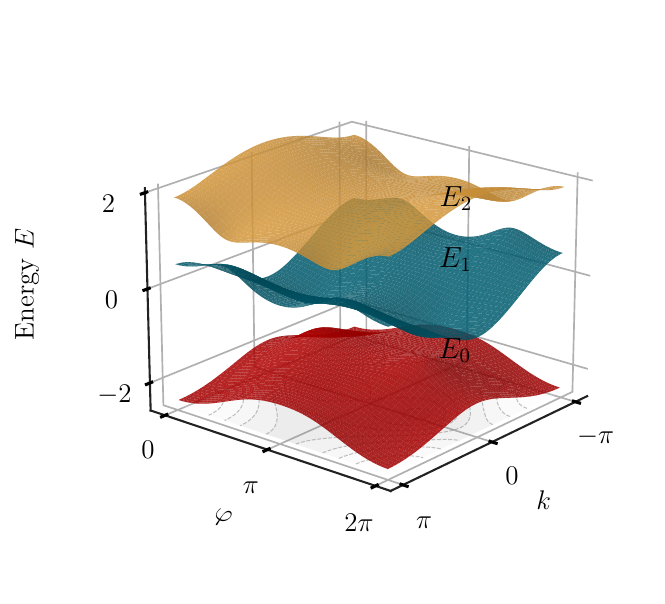}
    \caption{Three-dimensional instantaneous band structure of the trimer Thouless pump for $t_{0}=1$, $\delta_t=0.5$, and $M=0.5$. The three Bloch bands are plotted as surfaces over the $(k,\varphi)$ torus with $k\in[-\pi,\pi]$ and $\varphi\in[0,2\pi]$, and the axis ticks are expressed in units of $\pi$. For these parameters both bulk gaps remain finite throughout the cycle, and each band traces a smooth surface on the torus without closing. This regime realizes a fully adiabatic pump in which the global topology is encoded in the Berry curvature distributed over the three-band structure, rather than in any local touching singularity.}
    \label{fig:TTP_BANDS}
\end{figure}

\begin{figure}[t]
    \centering
    \includegraphics[width=\linewidth]{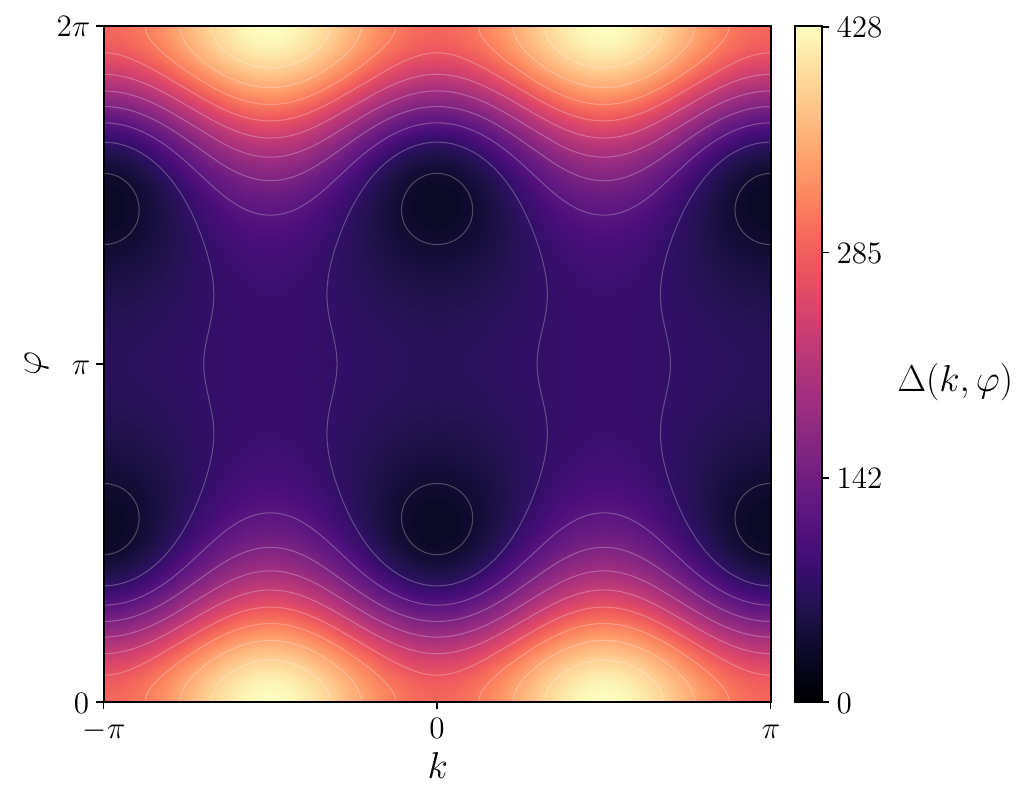}
    \caption{Discriminant $\Delta(k,\varphi)$ of the depressed cubic eigenvalue equation obtained from $\det[H(k,\varphi)-E\,\mathbb{I}]$ for the trimer Thouless pump with $t_{0}=1$, $\delta_t=0.5$, and $M=0.5$. The discriminant is evaluated over the $(k,\varphi)$ torus and plotted as a contour map using the dimensionless coordinates $k/\pi\in$ and $\varphi/\pi\in$. For a $3\times 3$ Hermitian Bloch Hamiltonian, $\Delta(k,\varphi)>0$ implies three real and nondegenerate eigenvalues at that point, whereas $\Delta=0$ would signal a band touching. In the present parameter regime $\Delta(k,\varphi)$ remains strictly positive and bounded away from zero across the entire torus, confirming that all three bands stay separated and that both bulk gaps remain open throughout the pumping cycle.}
    \label{fig:TTP_DISCRIMINANT}
\end{figure}

\subsubsection{Gap closings and Chern-number exchange}

The trimer contains three energy bands and two independent bulk gaps.
Topological phase transitions of the pump occur when one of these gaps
closes at isolated points on the $(k,\varphi)$ torus as system parameters
are tuned. For a fixed choice of hopping and onsite modulations, gap
closings are therefore codimension-one conditions in the enlarged space
spanned by $(k,\varphi)$ and the control parameters (such as $\delta_t$
and $M$).

The instantaneous band energies $E_n(k,\varphi)$ are the three real roots
of the cubic eigenvalue equation
\begin{equation}
\det\!\left[H(k,\varphi)-E\mathbb{I}\right]=0.
\end{equation}
For the TTP Hamiltonian \eqref{eq:Hkphi} with antisymmetric onsite
modulations $V_A(\varphi)=M\sin\varphi$ and $V_C(\varphi)=-M\sin\varphi$,
the trace
\begin{equation}
S_1 = V_A + V_C = 0
\end{equation}
vanishes identically. As a result, the characteristic polynomial is
already in depressed form and can be written as
\begin{equation}
E^{3} + S_{2}(k,\varphi)\,E - S_{3}(k,\varphi) = 0 ,
\label{eq:TTP_cubic_simplified}
\end{equation}
with
\begin{equation}
S_{2}
= V_{A}V_{C} - 2t_{1}^{2} - t^{2},
\qquad
S_{3}
= t_{1}^{2}\bigl(2t\cos k\bigr),
\end{equation}
where $t_1=t_1(\varphi)$, $t=t(\varphi)$, and $V_A=-V_C$ as in
Eqs.~\eqref{eq:tphi} and \eqref{eq:Vphi}. Equation
\eqref{eq:TTP_cubic_simplified} is thus of the standard depressed-cubic
form
\begin{equation}
\begin{split}
&E^{3} + p\,E + q = 0, \\
&p(k,\varphi)=S_{2}(k,\varphi), \quad q(k,\varphi)=-S_{3}(k,\varphi),
\end{split}
\end{equation}
Cardano's formula \cite{cardano2007rules} may be applied directly if desired.

The local eigenvalue structure is governed by the discriminant
\begin{equation}
\begin{split}
\Delta(k,\varphi) &= -4p(k,\varphi)^{3} - 27q(k,\varphi)^{2} \\
                  &= -4S_{2}^{3}(k,\varphi) - 27S_{3}^{2}(k,\varphi).
\end{split}
\label{eq:TTP_disc_simplified}
\end{equation}
For a $3\times3$ Hermitian Bloch Hamiltonian, the following statements
are exact:
\begin{itemize}
\item $\Delta(k,\varphi)>0$ implies three distinct real eigenvalues and
      both bulk gaps open at that point;
\item $\Delta(k,\varphi)=0$ indicates that at least two eigenvalues
      coincide, so that a bulk gap closes at that $(k,\varphi)$.
\end{itemize}
When the system is tuned to a critical point in parameter space where
$\Delta(k,\varphi)$ first vanishes at an isolated point
$(k^\ast,\varphi^\ast)$, two bands touch and exchange an integer unit of
Chern number. In the $(k,\varphi)$ description this band-touching point
acts as a Berry monopole whose quantized flux redistributes Berry
curvature between the two adjacent bands. Because the TTP has two
independent gaps, there can in principle be multiple such monopoles
associated with different band pairs, leading to a genuinely multigap
Chern-number redistribution that has no analogue in a two-band Thouless
pump.

For the specific parameter choice $t_0=1$, $\delta_t=0.5$, and $M=0.5$
used in Figs.~\ref{fig:TTP_BANDS} and \ref{fig:TTP_DISCRIMINANT}, our
analytic evaluation of $\Delta(k,\varphi)$ shows that it remains strictly
positive and bounded away from zero over the entire $(k,\varphi)$ torus.
In this regime all three bands are real and nondegenerate at every
$(k,\varphi)$, both bulk gaps stay open throughout the pumping cycle, and
the quantized transport is purely a consequence of the global Berry
curvature distribution rather than any local band-touching singularity.

\section{Berry Curvature Distribution, Band Topology, and Quantized Pumping}
\begin{figure*}[t]
    \centering
    \includegraphics[width=\textwidth]{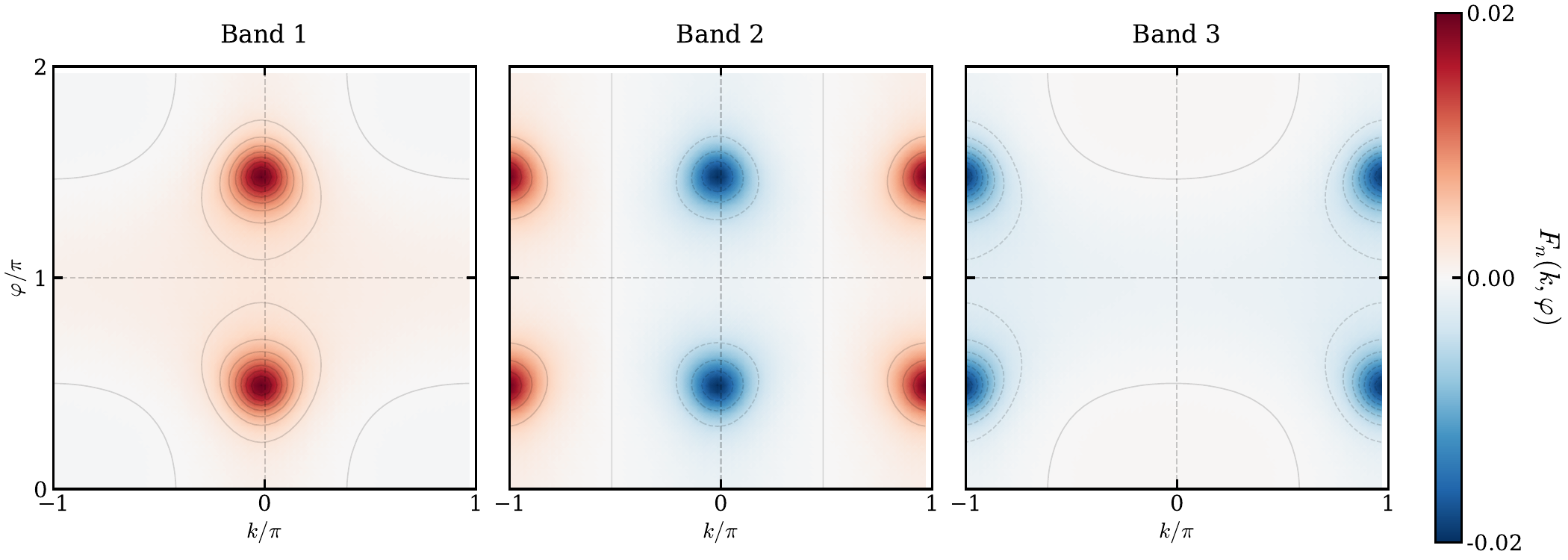}
    \caption{
    Berry curvature $F_n(k,\varphi)$ for all three bands of the
    trimer Thouless pump with $t_0 = 1$, $\delta_t = 0.5$, and $M = 0.5$.
    The curvature is plotted on the dimensionless torus
    $(k/\pi,\varphi/\pi)\in\times$ using the
    Fukui--Hatsugai--Suzuki discretization. Curvature is sharply localized
    at symmetry-related avoided crossings: for the lowest band near
    $(k,\varphi)\!\approx\!(0,\pi/2)$ and $(0,3\pi/2)$, for the highest band near
    $(\pm\pi,\pi/2)$ and $(\pm\pi,3\pi/2)$, and for the middle band in both
    regions with opposite signs. Integrating the curvature yields the
    quantized Chern numbers $(C_0,C_1,C_2)=(+1,0,-1)$, showing that the
    middle band mediates Berry-flux exchange between the two gaps while only
    the outer bands contribute net transported charge.
    }
    \label{fig:TTP_BC_HEATMAPS}
\end{figure*}
\label{sec:berry-curvature} The trimer Thouless pump supports three nondegenerate Bloch bands on the $(k,\varphi)$ torus. For each band $n$ we computed the Berry curvature\cite{berry1984quantal,zak1989berry} $F_n(k,\varphi)$ using the gauge-invariant discretization scheme introduced by Fukui, Hatsugai, and Suzuki\cite{Fukui2005}. The Berry curvature is defined on the two-dimensional parameter space spanned by the crystal momentum $k$ and the pump phase $\varphi$, and measures the geometric twisting of the occupied eigenstate under infinitesimal parallel transport on the torus. If a single band is filled, the quantized charge transported over one full adiabatic cycle is given by the band Chern number \begin{equation} C_n=\frac{1}{2\pi}\int_{-\pi}^{\pi} dk\int_{0}^{2\pi} d\varphi ~ F_n(k,\varphi), \label{eq:chern} \end{equation} and equals the net center-of-mass shift measured in units of the lattice spacing. This topological quantization persists in the presence of disorder and weak interactions as long as the bulk gap remains open during the cycle. We visualize the geometric structure of the pumping mechanism by plotting $F_n(k,\varphi)$ as a heatmap on the dimensionless torus $(k/\pi,\varphi/\pi)\in[-1,1]\times[0,2]$ for all three bands. A generic feature of the trimer pump is that Berry curvature is highly nonuniform: it is sharply localized near symmetry-enforced avoided crossings where two bands approach each other most closely. Away from those regions $F_n(k,\varphi)$ is effectively zero, indicating that most points of the torus contribute negligibly to the adiabatic polarization transport. For the representative parameters $t_{0}=1$, $\delta_t=0.5$, and $M=0.5$, the lowest band ($n=0$) develops two narrow curvature hotspots near $(k,\varphi)\approx(0,\pi/2)$ and $(0,3\pi/2)$. These peaks correspond to the strongest lower–middle hybridization events during the adiabatic cycle. Integrating $F_0(k,\varphi)$ over the torus yields the quantized value $C_0=+1$, showing that the lower band pumps exactly one unit of charge per cycle. The highest band ($n=2$) displays analogous curvature localization near $(k,\varphi)\approx(\pm\pi,\pi/2)$ and $(\pm\pi,3\pi/2)$. The sign of the curvature is opposite to that of the lowest band, and integration yields $C_2=-1$. Thus the upper and lower bands exchange a single quantum of Berry flux mediated by instantaneous mixing with the middle band. The middle band ($n=1$) develops curvature contributions near both sets of symmetry points—one inherited from the avoided crossing with the lower band, the other inherited from its avoided crossing with the upper band. These contributions appear with opposite signs and nearly cancel when summed over the whole torus, giving $C_1=0$. Although topologically trivial, the middle band remains geometrically active: it serves as the flux-exchange channel between the two gaps and mediates quantized pumping between the upper and lower subspaces. The Berry-curvature heatmaps thus provide a transparent picture of multiband charge pumping. Rather than being uniformly distributed, the curvature is concentrated in a handful of sharply localized regions on the torus. The quantized pumping does not rely on global uniformity but instead arises by integrating a small number of narrowly peaked geometric contributions. The integer relation \begin{equation} C_0 + C_1 + C_2 = 0 \end{equation} holds identically for any three-band Hermitian Bloch Hamiltonian, ensuring that the net geometric twist across the full Hilbert space vanishes. Viewed physically, the trimer Thouless pump implements polarization transfer through two avoided crossings located at different values of $k$. The topology of the lowest and highest bands is established by the directed adiabatic loop in parameter space, while the middle band acts as a dynamically coupled spectator that enables the redistribution of Berry curvature between gaps. This mechanism generalizes the Rice–Mele pump and represents the minimal example where geometric pumping is not encoded in a single two-band subspace but instead distributed across multiple gaps of a multiband lattice.

\section{Quantized Charge Pumping and Polarization Flow}
\label{sec:pumping}

The geometric structure revealed by the Berry-curvature distribution
admits a direct physical interpretation in terms of quantized charge
transport. In an adiabatic pump, the net charge transported across the
system during one full modulation cycle is equal to the Chern number of
the occupied Bloch bands. Although the instantaneous Hamiltonian remains
one-dimensional at every value of the pump phase $\varphi$, the combined
parameter space $(k,\varphi)$ endows the system with an effective
two-dimensional topology that governs the pumping response.

For the trimer Thouless pump, this translates to a genuinely band-resolved 
pumping mechanism. The lowest and highest bands drive quantized transport 
in opposite directions ($C_0=+1$, $C_2=-1$), while the middle band yields 
zero net transport ($C_1=0$). 

The Berry-curvature heatmaps clarify how this quantized response emerges.
Rather than being distributed uniformly over the $(k,\varphi)$ torus,
the curvature is sharply concentrated near a small number of
symmetry-related avoided crossings. These regions correspond to values of
$\varphi$ at which the onsite modulation is maximal and the instantaneous
band gaps are smallest, leading to enhanced interband mixing. Away from
these hotspots, the Berry curvature is negligible, indicating
that most of the adiabatic cycle does not contribute to charge transfer.

This localization implies that the pumping process proceeds through a
sequence of discrete polarization-transfer events rather than through a
continuous drift. As the pump parameter $\varphi$ passes through the
avoided crossings, the electronic polarization undergoes rapid changes 
associated with the localized exchange of geometric phase between 
neighboring bands. The total transported charge is obtained by
integrating these sharply localized contributions over the full cycle.

Importantly, the quantized pumping persists even though no bulk gap
closes during the cycle for the parameter regime considered here. The
topological transport is therefore protected by the global structure of
the adiabatic loop in parameter space rather than by any single
instantaneous band touching. This demonstrates that multiband Thouless
pumping can arise from the redistribution of Berry curvature across
multiple avoided crossings, providing a minimal realization of
quantized charge transport beyond the two-band Rice--Mele paradigm.

\section{Edge-State Spectral Flow and Localization}
\label{sec:edge-spectral-flow}
\begin{figure}[t]
    \centering
    \includegraphics[width=1.15\linewidth]{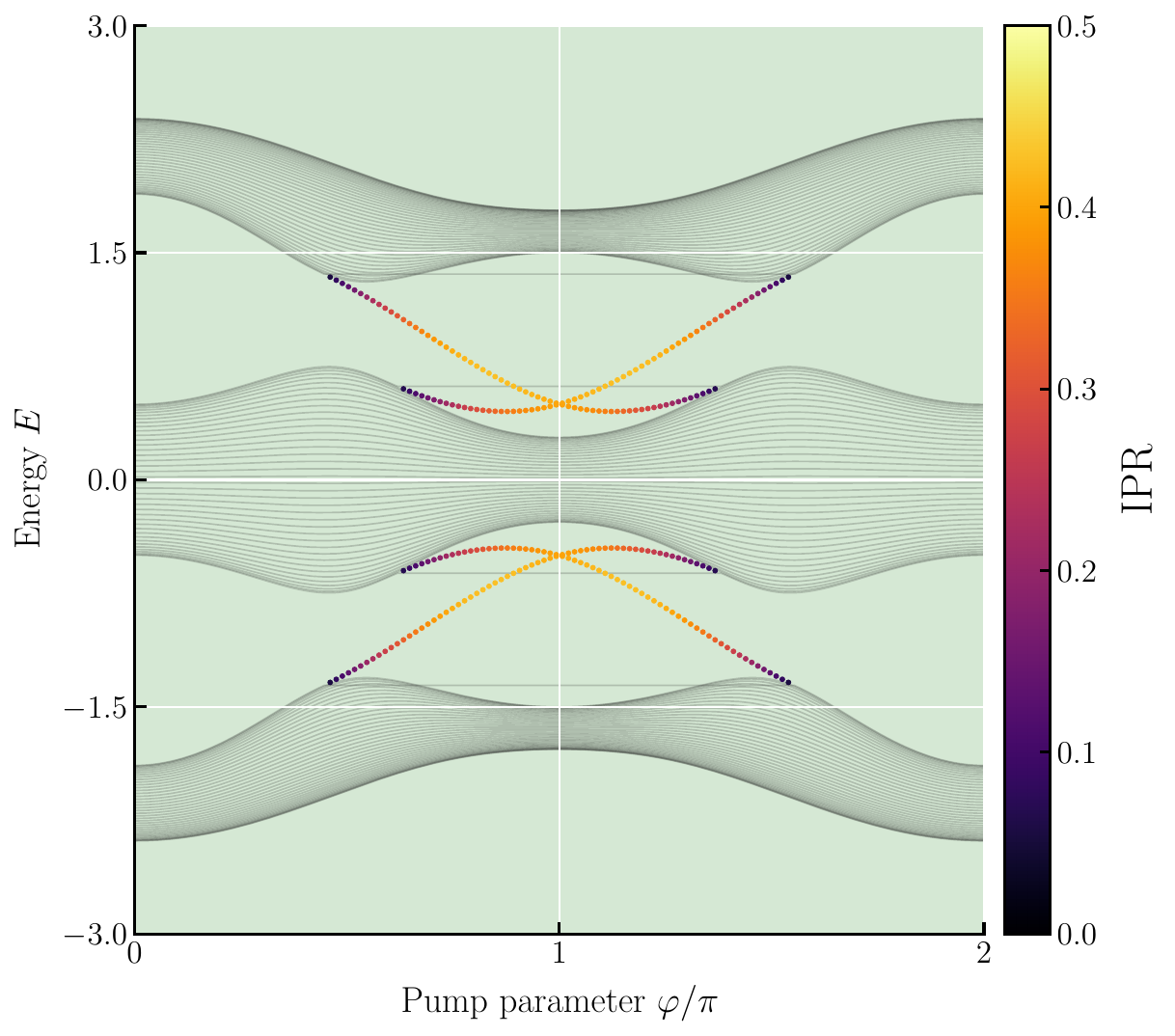}
\caption{
Instantaneous single-particle energy spectrum of the Trimer Thouless Pump with open boundary conditions as a function of the pump parameter $\varphi/\pi$ for $t_0=1$, $\delta_t=0.5$, $M=0.5$, and $N_{\mathrm{cell}}=40$. Each eigenstate is color-coded by its inverse participation ratio (IPR), distinguishing extended bulk states (low IPR) from boundary-localized edge states (high IPR). Bulk states form dense bands with weak localization, while highly localized edge states traverse the bulk gaps and connect different bands as $\varphi$ evolves. The direction and number of the gap-crossing edge branches are consistent with the bulk Chern numbers, demonstrating bulk--boundary correspondence for multigap Thouless pumping.
}
    \label{fig:TTP_edge_spectral_flow}
\end{figure}

Bulk topological pumping is accompanied by characteristic boundary
phenomena when the system is terminated with open boundary conditions.
To establish the bulk--boundary correspondence of the trimer Thouless
pump, we analyze the instantaneous energy spectrum of a finite open chain
as a function of the pump parameter $\varphi$ and identify boundary-localized
states via their inverse participation ratio (IPR).

We consider a chain of $N_{\mathrm{cell}}$ unit cells with open boundaries.
For each value of $\varphi$ the real-space Hamiltonian is diagonalized,
yielding the full set of instantaneous single-particle eigenenergies
$E(\varphi)$ and eigenstates.
Because translational symmetry is broken, crystal momentum is no longer
a good quantum number and the spectrum consists of $3N_{\mathrm{cell}}$
discrete levels.

To distinguish bulk-like states from boundary-localized states, we compute
the inverse participation ratio,
\begin{equation}
\mathrm{IPR} = \sum_j |\psi_j|^4 ,
\end{equation}
where $\psi_j$ denotes the normalized real-space wavefunction amplitude
on lattice site $j$.
Extended bulk states scale as $\mathrm{IPR}\sim 1/L$ and therefore appear
with small IPR, whereas states localized near the system boundaries exhibit
significantly enhanced IPR.

Figure~\ref{fig:TTP_edge_spectral_flow} shows the instantaneous energy
spectrum of the open chain as a function of the pump parameter $\varphi$,
with each eigenstate color-coded by its IPR.
Most eigenstates form dense bands with low IPR, corresponding to bulk
states that remain extended throughout the pumping cycle.
In contrast, a small number of states with large IPR emerge inside the
bulk gaps and trace continuous trajectories connecting different bulk
bands as $\varphi$ is varied.

These highly localized states represent topological edge modes whose
spectral flow reflects the nontrivial bulk topology.
The number and direction of gap-crossing edge-state branches are
consistent with the Chern numbers extracted from the bulk :
a net upward spectral flow across a gap corresponds to a positive Chern
number of the lower band, while a downward flow corresponds to a negative
Chern number.
The middle band, which carries zero Chern number, does not support a net
spectral flow but instead mediates the transfer of Berry curvature between
the upper and lower bands.

The coexistence of quantized bulk Chern numbers, sharply localized Berry
curvature hotspots, and boundary-localized spectral flow provides a
complete and internally consistent picture of multiband Thouless pumping
in the trimer lattice.
Unlike the two-band Rice--Mele pump, the present system exhibits edge
states associated with multiple bulk gaps, confirming the genuinely
multiband character of the pumping mechanism.

\section{Robustness under central-site detuning}

To test the stability of the multigap topology, we introduce an onsite
potential on the central sublattice,
\begin{equation}
H_{\mathrm{pert}} = V_B \sum_j b_j^\dagger b_j ,
\end{equation}
which shifts the energy of the $B$ site while keeping the hopping structure
and the adiabatic driving protocol unchanged. This perturbation provides a
direct way to examine how the topological pumping responds to changes in the
relative onsite energies.
\subsection{Exact Analytical Phase Boundaries}
Topological phase transitions in the adiabatic pump occur when a bulk energy gap closes at an isolated point $(k^*, \varphi^*)$ on the parameter torus. Owing to inversion symmetry at $\sin\varphi = 0$, the gap-closing events associated with the central-site detuning $V_B$ are anchored at high-symmetry points, specifically $(k^*, \varphi^*) = (\pi, \pi)$ (or $(0, \pi)$).

At $(k, \varphi) = (\pi, \pi)$, the antisymmetric onsite potentials vanish ($V_A(\pi) = V_C(\pi) = 0$), $t_1 \equiv t_0 - \delta_t$, and $t \equiv t_0 + \delta_t$. The $3 \times 3$ Bloch Hamiltonian simplifies to:
\begin{equation}
H(\pi, \pi) = \begin{pmatrix}
0 & t_1 & -t \\
t_1 & V_B & t_1 \\
-t & t_1 & 0
\end{pmatrix}.
\end{equation}

By transforming to an orthonormal inversion-symmetry basis $\{\mathbf{v}_1, \mathbf{v}_2, \mathbf{v}_3\}$, defined by:
\begin{align}
\mathbf{v}_1 &= \frac{1}{\sqrt{2}}\begin{pmatrix} 1 \\ 0 \\ -1 \end{pmatrix}, \quad 
\mathbf{v}_2 = \frac{1}{\sqrt{2}}\begin{pmatrix} 1 \\ 0 \\ 1 \end{pmatrix}, \nonumber \\
\mathbf{v}_3 &= \begin{pmatrix} 0 \\ 1 \\ 0 \end{pmatrix},
\end{align}
the odd-parity state $\mathbf{v}_1$ completely decouples with an exact eigenvalue:
\begin{equation}
E_1 = t = t_0 + \delta_t.
\end{equation}

The remaining even-parity subspace $\{\mathbf{v}_2, \mathbf{v}_3\}$ forms the reduced $2 \times 2$ block:
\begin{equation}
H_{\text{sub}} = \begin{pmatrix}
-t & \sqrt{2}t_1 \\
\sqrt{2}t_1 & V_B
\end{pmatrix},
\end{equation}
which yields the eigenenergies:
\begin{equation}
E_{\pm} = \frac{(V_B - t) \pm \sqrt{(V_B + t)^2 + 8t_1^2}}{2}.
\end{equation}

Equating the decoupled state energy to $E_+$ ($E_1 = E_+$) determines the gap-closing condition:
\begin{align}
t &= \frac{(V_B - t) + \sqrt{(V_B + t)^2 + 8t_1^2}}{2} \nonumber \\
&\implies 3t - V_B = \sqrt{(V_B + t)^2 + 8t_1^2}.
\end{align}
Squaring both sides and simplifying yields:
\begin{align}
(3t - V_B)^2 &= (V_B + t)^2 + 8t_1^2 \nonumber \\
&\implies t(t - V_B) = t_1^2 \nonumber \\
&\implies V_B^c = t - \frac{t_1^2}{t}.
\end{align}

Substituting $t = t_0 + \delta_t$ and $t_1 = t_0 - \delta_t$, we arrive at the closed-form topological phase transition boundaries:
\begin{equation}
V_B^c = \pm \frac{4 t_0 \delta_t}{t_0 + \delta_t}.
\label{eq:vbc_exact}
\end{equation}
For $t_0 = 1.0$ and $\delta_t = 0.5$, Eq.~(\ref{eq:vbc_exact}) yields $V_B^c = \pm 4/3 \approx \pm 1.333$, in exact agreement with the numerical phase transitions shown in Fig.~\ref{fig:chernVB}.

\subsection{Moderate detuning}

For moderate values of $V_B$ (approximately $-\frac{4}{3} \lesssim V_B \lesssim \frac{4}{3}$),
the band structure remains gapped throughout the $(k,\varphi)$ parameter space.
Although the Berry curvature is redistributed, its integral over the torus
is unchanged. The Chern numbers therefore remain quantized as
\begin{equation}
(C_0,C_1,C_2)=(+1,\,0,\,-1),
\end{equation}
demonstrating that the pumping mechanism is robust in this regime. The effect
of the detuning is limited to a smooth deformation of the band dispersion
without altering the topological character of the bands.

\subsection{Large negative detuning}

When $V_B$ is decreased further, the Chern numbers change to
\begin{equation}
(C_0,C_1,C_2)=(0,\,+1,\,-1).
\end{equation}
In this regime, the lowest band becomes topologically trivial, while the
upper two bands carry opposite Chern numbers. Physically, lowering the
central-site energy shifts the pumping activity toward the upper bands.

\subsection{Large positive detuning}

For sufficiently large positive values of $V_B$, the Chern numbers instead
become
\begin{equation}
(C_0,C_1,C_2)=(+1,\,-1,\,0).
\end{equation}
Here the highest band is topologically trivial, and the pumping involves the
lower and middle bands. This regime corresponds to raising the central-site
energy relative to the outer sites.

\begin{figure}[t]
\centering
\includegraphics[width=\columnwidth]{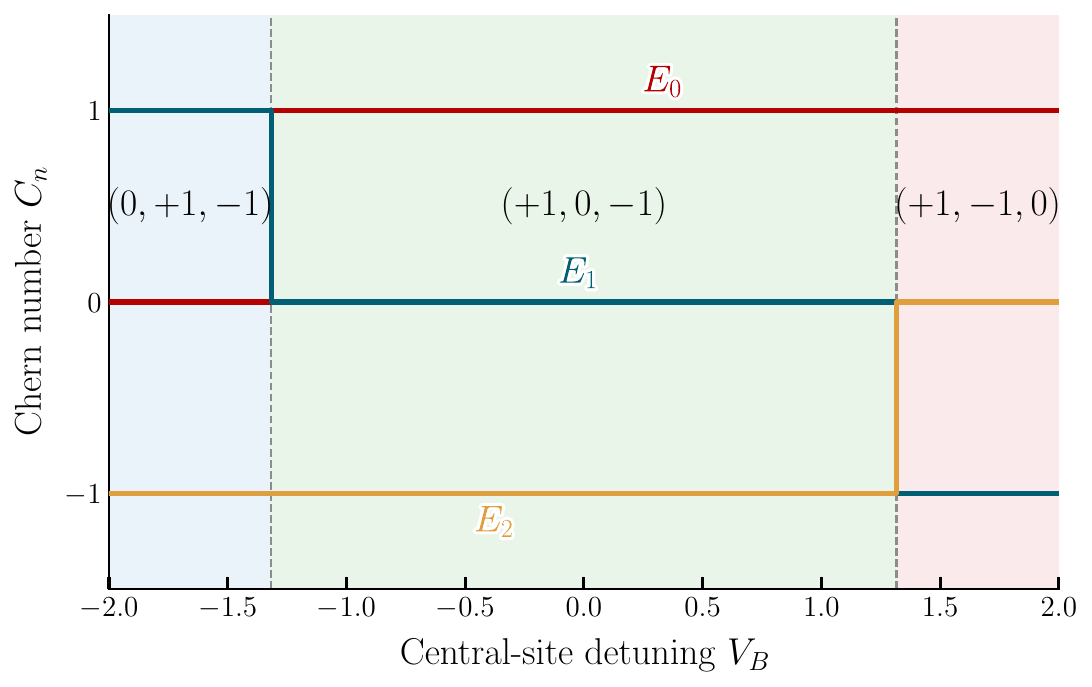}
\caption{
Band-resolved Chern numbers $C_n$ as functions of the central-site detuning $V_B$ for the Trimer Thouless Pump with $t_0=1$, $\delta_t=0.5$, and $M=0.5$. Distinct plateau regions correspond to the topological phases $(0,+1,-1)$, $(+1,0,-1)$, and $(+1,-1,0)$, while sharp changes between plateaus indicate topological phase transitions.
}
\label{fig:chernVB}
\end{figure}

\subsection{Summary}

The onsite detuning $V_B$ thus serves as a simple control parameter that
preserves the original pumping topology over a wide intermediate range and
induces controlled topological transitions when its magnitude becomes large.
\subsection{Band-selective transfer of Chern number}

The phase diagram obtained by varying the central-site detuning
reveals a simple organizing principle for multiband Thouless pumps.
Each topological transition occurs when an avoided crossing between
two neighboring bands reaches zero gap on the $(k,\varphi)$ torus.
The associated Berry curvature is localized near this crossing,
and the change in Chern number is confined to the two bands that
participate in the hybridization, while the third band remains
spectator-like.

From this perspective, the onsite detuning $V_B$ acts as a control
parameter that selects which pair of bands undergoes the effective
two-level Dirac dynamics discussed in Sec.\ref{sec:berry-curvature}. The redistribution of
Chern numbers observed in Fig.~\ref{fig:chernVB} therefore reflects a
band-selective transfer of Berry flux rather than the creation or
destruction of topological charge. This mechanism is not specific to
the present model but provides a minimal illustration of how topology
can be continuously rearranged within a multiband adiabatic pump.

\section{Experimental Realization in ultracold atoms}

A feasible experimental realization of the proposed trimer Thouless pump can be achieved using ultracold atoms loaded into a phase-controlled optical superlattice \cite{nakajima2016topological,lohse2016thouless}. This spatial imaging and extraction of boundary-localized transport takes structural cues from pioneering implementations of topological boundary pumping in specialized engineered lattices \cite{stutzer2018photonic}. Beyond these configurations, the tight-binding structure of the TTP can also be mapped onto solid-state simulators, drawing direct inspiration from foundational methods used for imaging topological edge states in silicon photonic platforms \cite{hafezi2013imaging}. The essential requirement is to generate an antisymmetric onsite potential modulation on the outer sublattices $A$ and $C$, while keeping the central sublattice $B$ at a stationary nodal point throughout the pumping cycle.

This can be implemented by superimposing a primary short optical lattice $V_s(x)$ defining the triple-well trimer geometry (with spatial period $3a$) and a phase-modulated long superlattice $V_l(x, \varphi)$. Choosing the central site $B$ of each unit cell as the spatial origin, the three sublattices are positioned at $x_A = -a$, $x_B = 0$, and $x_C = +a$.

The long-wavelength optical superlattice potential with wavevector $k_L = 2\pi/(3a)$ and tunable optical phase $\varphi$ is given by:
\begin{equation}
V_{\text{long}}(x, \varphi) = - V_L \sin\!\left(\frac{2\pi x}{3a}\right) \sin\varphi,
\label{eq:V_long}
\end{equation}
where $V_L$ is the optical potential depth, and $\varphi$ is varied adiabatically over a full pumping cycle $0 \le \varphi < 2\pi$.

Evaluating $V_{\text{long}}(x, \varphi)$ at the three sublattice positions yields:
\begin{subequations}
\begin{align}
V_A(\varphi) &= -V_L \sin\!\left(-\frac{2\pi}{3}\right)\sin\varphi = +\frac{\sqrt{3}}{2} V_L \sin\varphi, \\
V_B(\varphi) &= -V_L \sin(0) \sin\varphi = 0, \\
V_C(\varphi) &= -V_L \sin\!\left(+\frac{2\pi}{3}\right)\sin\varphi = -\frac{\sqrt{3}}{2} V_L \sin\varphi.
\end{align}
\label{eq:sublattice_potentials}
\end{subequations}

Setting $M = \frac{\sqrt{3}}{2} V_L$ reproduces exactly the antisymmetric onsite potential structure used in Eq.~(\ref{eq:Vphi}):
\begin{equation}
V_A(\varphi) = M \sin\varphi, \quad V_B(\varphi) = 0, \quad V_C(\varphi) = -M \sin\varphi.
\end{equation}
Crucially, because the central site $B$ coincides with the spatial node of $V_{\text{long}}(x, \varphi)$, it experiences zero onsite potential modulation throughout the entire adiabatic cycle.

Simultaneously, the modulated hoppings $t_1(\varphi) = t_0 + \delta_t \cos\varphi$ and $t(\varphi) = t_0 - \delta_t \cos\varphi$ can be engineered by dynamically controlling the relative intensity of a secondary lattice standing wave. Such phase-controlled superlattice schemes are well established in cold-atom experiments and represent a natural generalization of two-sublattice pumps to a three-sublattice geometry \cite{nakajima2016topological,lohse2016thouless,flaschner2016experimental}.

The quantized charge transport and band Chern numbers can be directly probed in cold-atom experiments by measuring the center-of-mass displacement of the atomic cloud after one full pumping cycle using in-situ absorption imaging \cite{nakajima2016topological,lohse2016thouless,flaschner2016experimental}.

\section{ SUMMARY AND DISCUSSION}

In this work, the Trimer Thouless Pump (TTP) was introduced and analyzed as a minimal three-band extension of the Rice--Mele model that provides a simple platform for investigating multigap topological charge pumping. By cyclically modulating the hopping amplitudes and antisymmetric onsite potentials, quantized charge transport was demonstrated across two independent bulk gaps characterized by the band Chern numbers $(+1,\,0,\,-1)$. Unlike conventional two-band pumps, the trimer lattice supports a richer redistribution of Berry curvature over the parameter torus, leading to distinct pumping channels associated with neighboring bulk gaps.

A central finding of this analysis concerns the distinctive geometric role of the middle energy band. This band acts as a geometric mediator and can be interpreted as mediating successive transfers of one unit of Chern number (equivalently, one quantum of Berry flux) between adjacent bands while maintaining a net zero Chern number itself. Because the Berry curvature is highly localized near isolated avoided crossings, the dominant topological response originates from localized regions of parameter space. This localized interpretation provides a physically transparent connection between the three-band trimer pump and local effective two-level Dirac descriptions associated with isolated avoided crossings, showing how the global Chern numbers can be understood as arising from successive adjacent-band topological exchanges. This bulk topological framework is further corroborated by the spectral flow of open-chain boundary states, where highly localized edge modes traverse multiple bulk gaps in directions dictated by the band-resolved Chern numbers.

Furthermore, by mapping the topological phase diagram as a function of the central-site detuning $V_B$, it was shown how multiband systems accommodate a band-selective redistribution of Chern numbers. These phase transitions demonstrate that the topology of the TTP is not rigidly fixed but can be continuously rearranged across different bulk gaps through the tuning of local site energies. In addition, a concrete experimental protocol was proposed for realizing this multigap transport using ultracold atoms in phase-controlled optical superlattices. By interfering two standing-wave optical potentials to generate an effective antisymmetric onsite modulation, the theoretical model is compatible with existing experimental capabilities. The resulting quantized pumping may be directly observed through measurements of the center-of-mass displacement of the atomic cloud.

The present work also suggests several natural directions for future investigation. While the analysis presented here focuses on the adiabatic regime, extending the model to nonadiabatic Floquet driving may reveal anomalous pumping channels and boundary states beyond those characterized by instantaneous Chern numbers. Incorporating many-body interactions into the trimer lattice may further lead to fractional charge pumping or interaction-induced topological phase transitions, thereby extending the present framework toward correlated topological transport. Moreover, generalizing the three-band architecture to two-dimensional lattices may provide insight into higher-order topological phases supporting corner or hinge states that are pumped across multiple bulk gaps.

The Trimer Thouless Pump therefore provides a minimal platform for investigating multigap topological transport and offers a useful starting point for exploring more complex multiband topological phenomena in synthetic quantum systems. More generally, the localized adjacent-band interpretation developed in this work provides a physically transparent conceptual picture for understanding how multiband topological pumps redistribute Berry curvature among neighboring bands while respecting the global topological constraint
\begin{equation}
\sum_{n} C_n = 0,
\end{equation}
thereby connecting multiband topological transport to the familiar physics of local effective two-level Dirac sectors associated with isolated avoided crossings.

\begin{acknowledgments}
R.C. acknowledges financial support from the University Grants Commission (UGC) fellowship (Award No.~211610042893).
\end{acknowledgments}
\section*{CRediT Author Statement}
\textbf{Rittwik Chatterjee:} Conceptualization, Methodology, Formal analysis, Investigation, Writing -- original draft, Visualization.

\section*{Declaration of Competing Interest}
The author declares no known competing financial interests or personal relationships that could have appeared to influence the work reported in this paper.

\section*{Data Availability}
The data that support the findings of this study are available from the corresponding author upon reasonable request.

\end{document}